%% file: main.tex
\documentclass[conference]{IEEEtran}
\IEEEoverridecommandlockouts
\usepackage{cite}
\usepackage{amsmath,amssymb,amsfonts}
\usepackage{algorithmic}
\usepackage{graphicx}
\usepackage{textcomp}
\usepackage{xcolor}
\usepackage{booktabs,balance}

\usepackage{booktabs}   
\usepackage{makecell}
\def\BibTeX{{\rm B\kern-.05em{\sc i\kern-.025em b}\kern-.08em
    T\kern-.1667em\lower.7ex\hbox{E}\kern-.125emX}}
\begin{document}

\title{SPEC CPU2026 Under the Microscope: Programs, Phases, and Samples}

\author{\IEEEauthorblockN{Peijun Li}
\IEEEauthorblockA{\textit{Computer Science and Engineering} \\
\textit{University of Minnesota, Twin Cities}\\
Minneapolis, MN, USA \\
li004367@umn.edu}
\and
\IEEEauthorblockN{Pei-Hung Lin}
\IEEEauthorblockA{\textit{Center for Applied Scientific Computing} \\
\textit{Lawrence Livermore National Laboratory}\\
Livermore, CA, USA \\
lin32@llnl.gov}
\and
\IEEEauthorblockN{Tianming Cui}
\IEEEauthorblockA{\textit{Computer Science and Engineering} \\
\textit{University of Minnesota, Twin Cities}\\
Minneapolis, MN, USA \\
cuixx327@umn.edu}
\and
\IEEEauthorblockN{Caiwen Ding}
\IEEEauthorblockA{\textit{Computer Science and Engineering} \\
\textit{University of Minnesota, Twin Cities}\\
Minneapolis, MN, USA \\
dingc@umn.edu}
\and
\IEEEauthorblockN{Yang Katie Zhao}
\IEEEauthorblockA{\textit{Electrical and Computer Engineering} \\
\textit{University of Minnesota, Twin Cities}\\
Minneapolis, MN, USA \\
yangzhao@umn.edu}
\and
\IEEEauthorblockN{Maya Gokhale}
\IEEEauthorblockA{\textit{Center for Applied Scientific Computing} \\
\textit{Lawrence Livermore National Laboratory}\\
Livermore, CA, USA \\
gokhale2@llnl.gov}
}

\maketitle

\begin{abstract}
\input{sections/0_abstract}
\end{abstract}

\begin{IEEEkeywords}
SPEC CPU2026, workload characterization, program behavior, sampled simulation, performance analysis
\end{IEEEkeywords}

\input{sections/1_introduction}
\input{sections/2_background}
\input{sections/3_methodology}

\input{sections/4_program}
\input{sections/5_phase}
\input{sections/6_sampling}
\input{sections/7_conclusion}

\input{sections/8_acknowledgement}

\balance
\bibliographystyle{IEEEtran}
\bibliography{refs}

\end{document}

%% file: sections/0_abstract.tex
SPEC CPU2026, the latest generation of the SPEC CPU benchmarks, is set to become the standard benchmark suite in computer architecture research, where new designs are often evaluated through sampled simulation with tools such as SimPoint. Existing characterizations of the suite focus on whole-run averages, which reveal neither how behavior changes within a run nor how well sampling captures it. We characterize CPU2026 on real hardware at three levels: programs, phases, and samples. At the program level, we report whole-run metrics and top-down bottleneck classes for all four suites and examine how they change with workload size, compiler version, and platform, across machines with three different instruction set architectures: x86, Arm, and RISC-V. Notably, changing only the compiler version can halve or nearly double the IPC of individual benchmarks. At the phase level, we track hardware counters every 10\,ms over all 90 SPECrate inputs, identify recurring patterns of time-varying behavior, and divide the inputs into three groups: steady, multi-phase, and bursty. More than half of the inputs show time-varying behavior, and IPC within a single input can vary by more than an order of magnitude. At the sample level, we validate SimPoint against hardware measurements of every interval it can select. SimPoint is accurate on steady inputs, but its error reaches 43\% on the others, and changing only the random seed spreads the error of one input over a range of 58 percentage points. Keeping SimPoint's clusters but choosing each representative by measured CPI brings every input within 2.6\%, so the error comes from the choice of representatives, not from the clustering. We conclude with practical recommendations on sampling strategy for CPU2026, which give researchers a deeper understanding of the suite and show where additional simulation effort is needed to keep sampled results accurate. 

%% file: sections/1_introduction.tex
\section{Introduction}
\label{sec:intro}
SPEC CPU has been the standard benchmark suite for evaluating processor performance for more than three decades~\cite{henning2000cpu2000, henning2006cpu2006, bucek2018cpu2017, spec2026method}, and each new release quickly becomes the default in architecture research~\cite{citron2003misspeculation, bera2021pythia, bai2023archexplorer, juffinger2024suit, niederer2025stackengine}. SPEC CPU2026, released in May 2026, replaces CPU2017 after nine years. It is not an incremental update: most of its benchmarks are new, and its source code is more than twice as large~\cite{spec2026press}. Alongside familiar workloads such as compilers, interpreters, and scientific codes, it adds applications from domains that have grown in importance since CPU2017, including electronic design automation, machine learning, graph analytics, and homomorphic encryption~\cite{spec2026method}. Workloads this different from their predecessors are likely to stress processors in new ways, so what we learned from CPU2017 cannot simply be carried over. As CPU2026 becomes the basis for evaluating processors, compilers, and simulators over the coming years, the community needs a detailed understanding of how it behaves.

Several characterization studies of CPU2026 have already appeared~\cite{spec2026method, li2026ut, kashyap2026amd, alotoom2026fidelity}. Most of them, like earlier studies of CPU2017~\cite{panda2018hpca, limaye2018ispass}, describe each benchmark with whole-run averages: a single IPC, a single miss rate per cache level, a single top-down breakdown. One of them~\cite{kashyap2026amd} even collects interval samples over time, but averages them into one value and discards periods that are not in steady state. Averages show which benchmarks stress which hardware structures, but they hide how behavior changes during a run: a benchmark that alternates between fast and slow phases can have the same average as one that runs at a constant speed.

This difference matters most in simulation. Architects evaluate new processor designs largely in cycle-level simulators such as gem5~\cite{binkert2011gem5}, which run several orders of magnitude slower than real hardware. At that speed, simulating a single full reference run of a CPU2026 benchmark would take weeks to months, and a design study must repeat this for every benchmark and every configuration it evaluates. In practice, researchers therefore simulate only a few short intervals and use them to estimate the behavior of the whole program, an approach known as sampled simulation~\cite{wunderlich2003smarts, sherwood2002simpoint}. SimPoint~\cite{sherwood2002simpoint, hamerly2005simpoint3} is the most widely used way to choose these intervals. It divides the execution into fixed-length intervals, clusters them by the code they execute, and selects one representative interval per cluster for simulation. This rests on an assumption: intervals that execute the same code also behave the same on the hardware~\cite{lau2005correlation}. Whether the assumption holds depends on exactly the time-varying behavior that whole-run averages hide. If it does not hold, a design evaluated on these intervals may look faster or slower than it really is.

This paper characterizes SPEC CPU2026 at three levels, programs, phases, and samples, each answering a question raised by the previous one. At the program level, we ask which hardware structures each benchmark stresses and how stable this picture is. We measure whole-run behavior for all four suites and all three workloads (test, train, and ref) on an Intel Sapphire Rapids (x86) machine and classify each benchmark by its dominant top-down bottleneck~\cite{yasin2014topdown}. We also rebuild the suites with another compiler version to measure how much the compiler alone changes the results, and, since the same source code compiles to different instructions on different instruction set architectures (ISAs), we run the SPECrate benchmarks on Arm and RISC-V machines and compare them with x86. These results guide the choice of benchmarks, but they are whole-run averages. At the phase level, we test whether such averages represent a run: we sample hardware performance counters every 10\,ms over full runs of the 90 SPECrate inputs and identify the inputs whose behavior changes over time. At the sample level, we ask how this behavior affects sampled simulation. We test SimPoint directly against real hardware: full runs are too long to simulate, so we instead measure every interval of each run on hardware and compare SimPoint's estimate with the whole-run measurement, which isolates the error introduced by sampling from any modeling error of a simulator. The phase classes then explain where SimPoint is accurate, and thus how much sampling effort each input needs.

This paper makes the following contributions:
\begin{itemize}
  \item \textbf{Program-level characterization.} Integer benchmarks are most often frontend bound and floating-point benchmarks memory bound. The train workload often matches the ref IPC but not its cache behavior, and changing only the compiler version can halve or nearly double the IPC of individual benchmarks.

  \item \textbf{Phase-level characterization.} Across the 90 SPECrate inputs on x86, whole-run averages hide large variation within a run. More than half of the inputs change behavior over time, and IPC within a single input can vary by more than an order of magnitude. We identify five patterns and classify the inputs as steady, multi-phase, or bursty.

  \item \textbf{Hardware validation of SimPoint.} SimPoint is accurate on steady inputs, but its error reaches 43\% on the others. Changing only the random seed spreads the error of one input over 58 percentage points, and a larger number of clusters does not fix the problem.

  \item \textbf{Root cause and recommendations.} Intervals that execute the same code can run at very different speeds. Keeping SimPoint's clusters but choosing each representative by measured CPI brings every input within 2.6\%, so the error comes from the choice of representatives, not the clustering. Based on these findings, we give practical recommendations for sampling CPU2026 with SimPoint.
  
\end{itemize}

The rest of this paper is organized as follows. Section~\ref{sec:bg} reviews SPEC CPU2026, prior characterization
studies, program phase behavior, and sampled simulation. Section~\ref{sec:method} describes our platforms and measurement methodology. Section~\ref{sec:program} presents the program-level results, Section~\ref{sec:phase} the phase-level results, and Section~\ref{sec:sampling} the validation of SimPoint and our recommendations. Section~\ref{sec:conclusion} concludes the paper.

%% file: sections/2_background.tex
\section{Background and Related Work}
\label{sec:bg}

\subsection{SPEC CPU2026}
\label{sec:bg-spec}
SPEC CPU2026 contains 52 benchmarks organized into four suites: SPECrate integer (14 benchmarks), SPECspeed integer (13), SPECrate floating point (12), and SPECspeed floating point (13)~\cite{spec2026method}. The Rate and Speed suites measure different things. SPECspeed measures how long one copy of each benchmark takes to complete, and many Speed benchmarks run in parallel, using OpenMP, C++ threads, or multiple processes. SPECrate measures throughput by running several single-threaded copies of each benchmark at the same time. Rate benchmarks are numbered 7xx with the suffix \texttt{\_r}, and Speed benchmarks 8xx with the suffix \texttt{\_s}. When an application appears in both suites, its two versions share a name and most of their source code but use different inputs, for example \texttt{727.cppcheck\_r} and \texttt{827.cppcheck\_s}.

Each benchmark comes with three workloads: test, train, and ref. Reported results use only ref; test is intended for quick correctness checks, and train for profile-guided optimization.

SPEC also modified many benchmarks to remove sources of non-determinism, for example by replacing \texttt{std::sort} with \texttt{std::stable\_sort}, which keeps equal elements in their original order. As a result, each benchmark performs the same work on every run, which makes its results reproducible.

\subsection{Characterization of SPEC CPU Suites}
\label{sec:bg-char}

Each new SPEC CPU release has been followed by characterization studies. For CPU2017, Panda et al.~\cite{panda2018hpca} measured hardware counters on several machines and used principal component analysis and hierarchical clustering to identify redundant benchmarks and select representative subsets. Limaye and Adegbija~\cite{limaye2018ispass} characterized all application-input pairs on a single machine and derived subsets from microarchitecture-independent features. Other studies focused on the memory system~\cite{singh2019memory,navarro2019memory} or on performance events and energy~\cite{hebbar2019}. These studies work at the level of whole benchmarks: each benchmark, or each application-input pair, is summarized by one value per metric. Subsetting reduces the number of benchmarks to simulate, but not the length of each one.

For CPU2026, the suite's developers describe how the benchmarks were selected and adapted, and report IPC and top-down breakdowns of the SPECrate benchmarks on an AMD Zen~5 processor~\cite{spec2026method}. They also show basic-block vector (BBV) recurrence plots and time-series performance plots for the single-threaded benchmarks, as a qualitative aid for selecting workloads. Li et al.~\cite{li2026ut} characterize the suite on nine platforms from Intel, AMD, Ampere, and NVIDIA, covering both x86 and Arm, and compare it with CPU2017, DCPerf, and MLPerf; they find higher instruction-cache and frontend pressure than in CPU2017. Kashyap et al.~\cite{kashyap2026amd} analyze the suite on AMD Zen~5 with top-down analysis and report one averaged value per benchmark. 

\subsection{Program Phases and Sampled Simulation}
\label{sec:bg-sampling}
Programs do not behave uniformly over their execution: their behavior changes over time and often repeats, a property known as phase behavior~\cite{sherwood2002simpoint,duesterwald2003}. Phases can be identified from the code a program executes, for example with BBVs, which record how often each basic block runs within an interval, or directly from hardware metrics such as IPC.

Sampled simulation methods fall into two families: statistical sampling, such as SMARTS~\cite{wunderlich2003smarts}, which simulates many very short samples spread uniformly across the execution, and phase-based sampling, such as SimPoint~\cite{sherwood2002simpoint,hamerly2005simpoint3}, which selects one representative interval per program phase for simulation. Multi-threaded programs require dedicated methods such as BarrierPoint~\cite{carlson2014barrierpoint} and LoopPoint~\cite{sabu2022looppoint}; our sampling study uses single-threaded SPECrate benchmarks and focuses on SimPoint.

SimPoint identifies phases by dividing the execution into fixed-length intervals and collecting a BBV for each. It clusters the BBVs with $k$-means after a random projection, choosing the number of clusters $K$ up to a user-specified maximum, and picks the interval closest to each cluster center as its representative, weighted by the cluster's share of intervals. Random seeds control both the projection and the initial cluster centers, so different seeds can yield different representatives.

SimPoint was originally validated against full detailed simulation on SPEC CPU2000~\cite{sherwood2002simpoint}, and simulation points have since been produced for newer suites~\cite{nair2008cpu2006,wu2018hot}. For CPU2017, Han and Hardavellas compared SimPoint-based checkpoints with native whole-program execution and reported an average CPI error of 12\%~\cite{han2021pinpoints}. Caculo et al.\ found that clusters formed from BBVs alone can mix low- and high-IPC intervals~\cite{caculo2025mav}, and SimPoint+ makes the results more stable across runs~\cite{xue2025simpointplus}. Unlike these works, we validate SimPoint on CPU2026 against per-interval hardware measurements for every input, and analyze where its error comes from.

%% file: sections/3_methodology.tex
\section{Methodology}
\label{sec:method}

\subsection{Experimental Setup}
\label{sec:method-setup}

Table~\ref{tab:platforms} lists our three platforms: an Intel Sapphire Rapids (x86) machine, a Graviton4 (Arm), and a SpacemiT K3 (RISC-V). The K3 combines eight X100 general-purpose cores with eight A100 cores designed for AI workloads; we use only the X100 cores. All platforms run Linux, and we read hardware counters with \texttt{perf}. We compile all benchmarks at \texttt{-O3} for the native microarchitecture of each platform, so that each platform uses its own vector extension, and with the same compiler version on all platforms. This removes the compiler as a source of difference between platforms; the remaining differences come from both the ISA and the microarchitecture, which our cross-platform results do not separate.

SPEC provides three workloads for each benchmark: test, train, and ref. The program-level study (Section~\ref{sec:program}) covers all four suites with all three workloads on the x86 platform, and both SPECrate suites with the ref workload on all three platforms; \texttt{766.femflow\_r} fails to compile on the RISC-V platform, so the RISC-V results cover 25 of the 26 Rate benchmarks. The phase-level and sampling studies (Sections~\ref{sec:phase} and~\ref{sec:sampling}) use the ref workload of the SPECrate suites on the x86 platform. In this workload, several of the 26 benchmarks run multiple times with different input files, giving 90 inputs in total (57 integer and 33 floating point). We name each input by its benchmark and position, for example \texttt{721.gcc\_r/3}.

We run each SPECrate benchmark as a single copy and each SPECspeed benchmark with a single thread, pinned to one core. All runs are launched through SPEC's \texttt{runcpu} harness, with our measurement tools attached through its monitoring hooks, so that every run goes through SPEC's output validation.

\begin{table}[t]
\centering\footnotesize
\caption{Evaluated platforms.}
\label{tab:platforms}
\setlength{\tabcolsep}{3pt}
\begin{tabular*}{\columnwidth}{@{\extracolsep{\fill}}llll}
\toprule
 & x86 & Arm & RISC-V \\
\midrule
Processor & Xeon 8480+ & Graviton4 & SpacemiT K3 \\
Core & Golden Cove & Neoverse V2 & X100 \\
Max freq. & 3.8\,GHz & 2.8\,GHz & 2.15\,GHz \\
L1I / L1D & 32 / 48\,KB & 64 / 64\,KB & 64 / 64\,KB \\
L2 & 2\,MB & 2\,MB & 4\,MB$^\dagger$ \\
L3 & 105\,MB & 36\,MB & -- \\
Memory & 256\,GB & 192\,GB & 32\,GB \\
Vector ext. & AVX-512 & SVE2 & RVV \\
Compiler & \makecell[l]{GCC 15.2.0\\GCC 13.3.1$^\ddagger$} & GCC 15.2.0 & GCC 15.2.0 \\
Flags & \makecell[l]{\texttt{-O3 -march=}\\\texttt{x86-64-v4}}
      & \makecell[l]{\texttt{-O3 -mcpu=}\\\texttt{native}}
      & \makecell[l]{\texttt{-O3 -mcpu=}\\\texttt{spacemit-x100}} \\
\bottomrule
\end{tabular*}
\par\smallskip\raggedright\scriptsize
$^\dagger$Shared by four cores.
$^\ddagger$Used only for the compiler study in Section~\ref{sec:compiler}.
\end{table}

\subsection{Hardware Measurement}
\label{sec:method-hw}

We measure hardware counters at three granularities, one for each level of our study. All counters count user-mode events only, matching the user-level instruction stream that Pin~\cite{luk2005pin} instruments for BBV collection.

\textit{Whole run.} For the program-level study, we count events over the entire run of each input; Section~\ref{sec:program} describes the metrics derived from them. On x86 we also collect the top-down breakdown~\cite{yasin2014topdown} using the support built into \texttt{perf}, which counts across all privilege levels.

\textit{Every 10\,ms.} For the phase-level study, we sample hardware counters every 10\,ms with \texttt{perf stat -I}, giving a time series of IPC and miss rates for each input.

\textit{Every 100 million instructions.} SimPoint defines intervals by instruction count, so validating it requires measurements on the same boundaries. We sample on the retired-instruction counter every 100 million instructions and record the cycle count at each sample, which gives the cycles of every interval. Since a program executes essentially the same instruction sequence for a given input each time it is run, these per-interval cycle counts serve as ground truth for the SimPoint validation.

\subsection{BBV Collection and SimPoint}
\label{sec:method-simpoint}

We collect BBVs on x86 with a Pin~4.4 tool, using intervals of 100 million user-mode instructions, a common SimPoint granularity and the same period as the hardware sampling. The tool's total instruction count agrees with the hardware counter to within 0.65\% for every input, and we map each BBV interval to the hardware interval at the same relative position in the run, which removes the remaining drift.

We cluster the BBVs with SimPoint~3.2~\cite{hamerly2005simpoint3} at its default settings, letting it choose the number of clusters automatically with an upper bound of 30, yielding one representative interval and its weight per cluster. Section~\ref{sec:sampling} examines the effect of the interval size and repeats the clustering with five random seeds and once with a fixed number of clusters $K=30$.

%% file: sections/4_program.tex
\section{Program-Level Characterization}
\label{sec:program}

\subsection{Whole-Run Metrics}

\begin{table*}[!t]
\centering\scriptsize
\caption{Whole-run characteristics of SPEC CPU2026 (ref workload, single copy/thread, Sapphire Rapids, GCC 15.2.0).}
\label{tab:wholerun}
\setlength{\tabcolsep}{3.5pt}
\begin{tabular}{llrrrrrrrrrrrrr}
\toprule
Benchmark & Application & \makecell{Instr.\\(G)} & IPC & \makecell{Load\\(\%)} & \makecell{Store\\(\%)} & \makecell{Branch\\(\%)} & \makecell{Br.\ mis-\\pred.\ (\%)} & \makecell{L1I\\MPKI} & \makecell{L1D\\MPKI} & \makecell{LLC\\MPKI} & \makecell{Front-\\end (\%)} & \makecell{Bad\\spec.\ (\%)} & \makecell{Retir-\\ing (\%)} & \makecell{Back-\\end (\%)} \\
\midrule
\multicolumn{15}{l}{\textit{SPECrate 2026 Integer} (14 benchmarks)} \\
\texttt{706.stockfish\_r} & Chess engine & 987 & 1.67 & 24.7 & 10.1 & 9.7 & 4.50 & 8.95 & 29.58 & 0.59 & 23.1 & 14.5 & 28.0 & 34.3 \\
\texttt{707.ntest\_r} & Game AI & 2,000 & 3.01 & 27.6 & 7.9 & 9.1 & 2.83 & 0.01 & 2.59 & 0.00 & 11.2 & 23.1 & 48.7 & 17.0 \\
\texttt{708.sqlite\_r} & SQL database & 1,692 & 2.48 & 29.5 & 12.2 & 20.9 & 0.52 & 1.00 & 2.78 & 1.40 & 15.6 & 11.2 & 38.4 & 34.9 \\
\texttt{710.omnetpp\_r} & Event simulation & 1,460 & 2.85 & 31.6 & 16.8 & 20.6 & 0.37 & 9.79 & 6.33 & 0.05 & 36.8 & 7.0 & 46.4 & 9.8 \\
\texttt{714.cpython\_r} & Interpreter & 1,422 & 3.65 & 27.4 & 15.4 & 21.5 & 0.08 & 23.57 & 0.23 & 0.06 & 39.4 & 1.8 & 56.7 & 2.2 \\
\texttt{721.gcc\_r} & C compiler & 1,060 & 1.54 & 27.7 & 11.2 & 22.4 & 1.54 & 11.80 & 13.49 & 3.20 & 26.5 & 11.2 & 23.1 & 39.1 \\
\texttt{723.llvm\_r} & C/C++ compiler & 954 & 1.41 & 24.1 & 13.1 & 21.3 & 2.87 & 22.69 & 6.99 & 3.18 & 38.2 & 16.3 & 21.4 & 24.1 \\
\texttt{727.cppcheck\_r} & Static analysis & 1,123 & 2.36 & 20.5 & 7.3 & 28.7 & 0.24 & 0.90 & 4.52 & 1.86 & 32.9 & 3.3 & 32.7 & 31.1 \\
\texttt{729.abc\_r} & Logic synthesis & 1,287 & 2.33 & 25.9 & 8.4 & 15.9 & 2.24 & 0.06 & 7.38 & 0.88 & 14.6 & 23.8 & 34.8 & 26.8 \\
\texttt{734.vpr\_r} & FPGA place \& route & 1,205 & 2.37 & 31.5 & 11.1 & 19.5 & 1.28 & 0.49 & 8.49 & 0.95 & 20.7 & 16.5 & 37.3 & 25.5 \\
\texttt{735.gem5\_r} & Arch. simulation & 1,153 & 2.30 & 31.1 & 14.3 & 21.5 & 0.90 & 16.95 & 15.87 & 0.06 & 36.8 & 11.1 & 36.4 & 15.7 \\
\texttt{750.sealcrypto\_r} & Homomorphic encryption & 2,390 & 4.56 & 12.5 & 5.6 & 3.4 & 0.51 & 0.01 & 0.75 & 0.05 & 3.1 & 2.3 & 75.9 & 18.7 \\
\texttt{753.ns3\_r} & Network simulation & 1,231 & 2.38 & 29.2 & 16.2 & 22.7 & 0.61 & 26.00 & 9.26 & 0.05 & 45.0 & 7.5 & 37.8 & 9.8 \\
\texttt{777.zstd\_r} & Data compression & 1,800 & 2.30 & 20.4 & 7.8 & 12.0 & 2.71 & 0.02 & 8.04 & 0.35 & 6.6 & 23.4 & 35.5 & 34.6 \\
\midrule
\multicolumn{15}{l}{\textit{SPECrate 2026 Floating Point} (12 benchmarks)} \\
\texttt{709.cactus\_r} & General relativity & 984 & 1.58 & 64.2 & 8.3 & 1.4 & 0.10 & 92.14 & 114.53 & 1.63 & 39.1 & 0.8 & 30.6 & 29.5 \\
\texttt{722.palm\_r} & Atmospheric science & 2,523 & 2.84 & 36.1 & 9.3 & 4.2 & 0.35 & 12.34 & 13.16 & 2.82 & 9.8 & 1.6 & 46.9 & 41.8 \\
\texttt{731.astcenc\_r} & Image compression & 1,855 & 2.19 & 27.8 & 5.3 & 9.4 & 7.16 & 1.88 & 1.81 & 0.00 & 20.7 & 28.1 & 37.2 & 14.0 \\
\texttt{736.ocio\_r} & Color management & 1,826 & 3.04 & 31.1 & 10.3 & 8.9 & 0.11 & 0.00 & 1.24 & 0.75 & 4.0 & 1.6 & 50.8 & 43.6 \\
\texttt{737.gmsh\_r} & Mesh generation & 940 & 1.68 & 29.7 & 11.8 & 17.1 & 1.96 & 0.33 & 7.29 & 1.31 & 16.6 & 18.6 & 26.8 & 38.0 \\
\texttt{748.flightdm\_r} & Flight dynamics & 1,716 & 2.94 & 29.3 & 13.4 & 18.7 & 0.06 & 18.29 & 18.18 & 0.00 & 18.4 & 2.0 & 48.1 & 31.6 \\
\texttt{749.fotonik3d\_r} & Electromagnetics & 494 & 0.66 & 42.3 & 12.7 & 3.1 & 0.08 & 0.02 & 57.73 & 38.61 & 2.4 & 0.4 & 12.9 & 84.3 \\
\texttt{765.roms\_r} & Ocean modeling & 1,022 & 0.97 & 38.1 & 9.8 & 10.3 & 0.06 & 0.08 & 49.40 & 17.01 & 2.7 & 0.8 & 17.9 & 78.6 \\
\texttt{766.femflow\_r} & CFD (FEM) & 1,145 & 2.71 & 36.0 & 15.5 & 4.6 & 0.28 & 0.87 & 14.63 & 0.80 & 3.5 & 1.2 & 44.4 & 51.0 \\
\texttt{767.nest\_r} & Neuroscience & 1,676 & 3.45 & 34.5 & 12.6 & 15.0 & 0.37 & 0.04 & 20.65 & 0.14 & 8.8 & 8.8 & 58.6 & 23.9 \\
\texttt{772.marian\_r} & Machine translation & 1,483 & 2.58 & 19.5 & 3.8 & 8.6 & 0.15 & 0.04 & 7.81 & 0.33 & 12.0 & 3.1 & 43.2 & 41.6 \\
\texttt{782.lbm\_r} & CFD (LBM) & 1,596 & 2.08 & 22.9 & 11.5 & 1.2 & 0.16 & 0.00 & 0.31 & 5.25 & 1.2 & 0.4 & 33.7 & 64.7 \\
\midrule
\multicolumn{15}{l}{\textit{SPECspeed 2026 Integer} (13 benchmarks)} \\
\texttt{801.xz\_s} & Data compression & 18,419 & 1.15 & 22.0 & 7.7 & 15.4 & 5.99 & 0.03 & 8.09 & 2.25 & 13.3 & 21.2 & 22.5 & 43.0 \\
\texttt{807.ntest\_s} & Game AI & 129,853 & 3.09 & 22.2 & 7.4 & 6.2 & 2.17 & 0.01 & 7.26 & 0.00 & 12.4 & 23.3 & 47.3 & 17.1 \\
\texttt{817.flac\_s} & Audio compression & 19,966 & 2.93 & 16.0 & 3.8 & 10.7 & 0.27 & 0.13 & 0.34 & 0.00 & 4.7 & 2.1 & 53.5 & 39.7 \\
\texttt{821.gcc\_s} & C compiler & 100,703 & 2.07 & 26.0 & 11.3 & 22.7 & 1.78 & 12.27 & 7.60 & 0.75 & 41.1 & 15.7 & 31.1 & 12.1 \\
\texttt{823.llvm\_s} & C/C++ compiler & 98,362 & 1.98 & 22.0 & 11.4 & 23.7 & 0.80 & 5.47 & 5.01 & 1.30 & 32.9 & 9.7 & 30.0 & 27.4 \\
\texttt{827.cppcheck\_s} & Static analysis & 86,182 & 2.48 & 22.3 & 9.7 & 27.3 & 0.32 & 2.58 & 4.93 & 1.37 & 35.8 & 4.9 & 35.0 & 24.3 \\
\texttt{829.abc\_s} & Logic synthesis & 1,163 & 0.88 & 24.7 & 10.7 & 17.8 & 1.74 & 0.09 & 12.28 & 7.55 & 9.3 & 11.2 & 14.1 & 65.4 \\
\texttt{834.vpr\_s} & FPGA place \& route & 2,651 & 2.38 & 32.0 & 11.1 & 20.2 & 1.18 & 0.34 & 8.40 & 1.04 & 25.9 & 7.2 & 40.4 & 26.5 \\
\texttt{835.gem5\_s} & Arch. simulation & 2,208 & 2.19 & 31.9 & 13.8 & 19.0 & 1.24 & 34.33 & 13.71 & 0.03 & 34.6 & 9.3 & 41.0 & 15.2 \\
\texttt{838.diamond\_s} & Metagenomics & 141,420 & 3.06 & 18.0 & 6.5 & 5.6 & 1.12 & 0.04 & 0.13 & 0.05 & 4.2 & 8.2 & 49.8 & 37.8 \\
\texttt{846.minizinc\_s} & Constraint solving & 4,462 & 2.57 & 27.0 & 18.2 & 15.1 & 0.90 & 2.18 & 7.97 & 0.08 & 16.4 & 8.9 & 39.3 & 35.4 \\
\texttt{853.ns3\_s} & Network simulation & 1,899 & 1.91 & 29.2 & 16.0 & 22.7 & 0.53 & 25.57 & 17.96 & 0.09 & 34.5 & 5.9 & 31.2 & 28.4 \\
\texttt{854.graph500\_s} & Graph analytics & 32,188 & 1.23 & 40.5 & 0.3 & 28.8 & 0.28 & 0.00 & 18.37 & 15.54 & 32.6 & 2.0 & 18.5 & 46.9 \\
\midrule
\multicolumn{15}{l}{\textit{SPECspeed 2026 Floating Point} (13 benchmarks)} \\
\texttt{800.pot3d\_s} & Solar physics & 2,071 & 0.35 & 32.9 & 8.3 & 6.9 & 0.01 & 0.01 & 119.06 & 70.43 & 1.9 & 0.4 & 7.4 & 90.3 \\
\texttt{803.sph\_exa\_s} & Astrophysics & 45,636 & 1.95 & 26.1 & 2.7 & 14.1 & 0.84 & 0.00 & 37.03 & 0.60 & 9.3 & 7.5 & 38.0 & 45.1 \\
\texttt{809.cactus\_s} & General relativity & 19,549 & 1.47 & 64.9 & 8.2 & 1.3 & 0.07 & 93.07 & 115.44 & 2.36 & 30.0 & 0.4 & 28.8 & 40.8 \\
\texttt{811.tealeaf\_s} & High-energy physics & 36,914 & 3.09 & 22.6 & 5.3 & 4.0 & 0.00 & 0.00 & 1.77 & 3.73 & 3.1 & 0.4 & 61.6 & 34.9 \\
\texttt{816.nab\_s} & Molecular modeling & 70,349 & 2.52 & 32.6 & 6.9 & 13.5 & 1.19 & 0.00 & 1.77 & 0.16 & 7.9 & 13.6 & 40.9 & 37.6 \\
\texttt{820.cloverleaf\_s} & Hydrodynamics & 16,452 & 1.67 & 37.7 & 4.1 & 6.8 & 0.01 & 0.00 & 15.39 & 10.09 & 5.0 & 1.2 & 30.6 & 63.2 \\
\texttt{822.palm\_s} & Atmospheric science & 34,863 & 2.82 & 36.2 & 10.3 & 3.9 & 0.28 & 4.87 & 12.31 & 2.26 & 5.2 & 0.8 & 41.3 & 52.8 \\
\texttt{849.fotonik3d\_s} & Electromagnetics & 13,969 & 1.81 & 61.1 & 8.2 & 2.7 & 0.16 & 0.00 & 11.31 & 23.31 & 6.7 & 1.9 & 33.0 & 58.4 \\
\texttt{857.namd\_s} & Molecular dynamics & 113,987 & 3.36 & 31.9 & 6.9 & 3.1 & 4.80 & 0.00 & 5.47 & 0.19 & 11.2 & 15.4 & 58.7 & 14.7 \\
\texttt{865.roms\_s} & Ocean modeling & 9,946 & 0.79 & 38.3 & 9.8 & 10.2 & 0.10 & 0.03 & 61.57 & 21.01 & 5.0 & 1.5 & 16.8 & 76.7 \\
\texttt{867.nest\_s} & Neuroscience & 63,738 & 0.87 & 30.2 & 10.5 & 13.4 & 1.42 & 0.02 & 19.69 & 9.31 & 12.9 & 6.7 & 37.2 & 43.2 \\
\texttt{872.marian\_s} & Machine translation & 14,284 & 1.90 & 21.3 & 6.4 & 6.5 & 0.08 & 0.02 & 9.87 & 0.54 & 7.4 & 2.1 & 37.6 & 53.0 \\
\texttt{881.neutron\_s} & Neutron transport & 33,494 & 1.05 & 25.3 & 9.8 & 10.4 & 2.51 & 0.01 & 11.82 & 8.56 & 9.9 & 18.6 & 16.5 & 55.0 \\
\bottomrule
\end{tabular}
\end{table*}

Table~\ref{tab:wholerun} summarizes the whole-run behavior of all 52 benchmarks on the ref workload on the x86 platform.

\textit{Scale.} Each Rate benchmark executes between 0.5 and 2.5 trillion instructions (494\,G for \texttt{749.fotonik3d\_r}, 2,523\,G for \texttt{722.palm\_r}). Speed benchmarks are far larger, up to 141 trillion for \texttt{838.diamond\_s}, although six of them execute fewer than 5 trillion. Even the smallest benchmark is far too long to simulate in full with a cycle-level simulator.

\textit{Instruction mix and branches.} Loads make up 20--40\% of the instructions in 43 of the 52 benchmarks. The two \texttt{cactus} benchmarks (64--65\%) and \texttt{849.fotonik3d\_s} (61\%) are dominated by loads, while \texttt{750.sealcrypto\_r} (13\%) sits at the other extreme. \texttt{854.graph500\_s}, which runs graph-search kernels over a large graph, issues almost no stores (0.3\%). Branch density separates integer from floating-point benchmarks: 21 of the 27 integer benchmarks have 12--29\% branches, whereas 22 of the 25 floating-point benchmarks have fewer than 15\%, so the integer suite exercises the branch predictor more often. The highest misprediction rates, however, include two floating-point benchmarks: \texttt{731.astcenc\_r} (7.2\%), \texttt{801.xz\_s} (6.0\%), and \texttt{857.namd\_s} (4.8\%).

\textit{Cache behavior.} Several benchmarks stand out in L1I misses per thousand instructions (MPKI), led by the two \texttt{cactus} benchmarks (92--93), \texttt{835.gem5\_s} (34), and the two \texttt{ns3} benchmarks (26). The highest last-level cache (LLC) MPKI belongs to \texttt{800.pot3d\_s} (70), which also has the lowest IPC in the suite (0.35), whereas five benchmarks, such as \texttt{707.ntest\_r}, have almost no LLC misses (below 0.005 MPKI). A high L1D miss rate alone does not imply heavy memory traffic: \texttt{709.cactus\_r} incurs 115 L1D misses per thousand instructions but only 1.6 LLC misses, so its working set fits in the lower-level caches.

\textit{Rate versus Speed.} Sharing a name does not imply sharing behavior. The \texttt{vpr} pair runs at the same IPC (2.37 and 2.38) with a similar instruction mix and similar miss rates, but \texttt{829.abc\_s} runs at an IPC of 0.88 with an LLC MPKI of 7.6, compared with 2.33 and 0.88 for \texttt{729.abc\_r}. Speed results therefore cannot be inferred from Rate results, or vice versa.

\subsection{Workload Scaling: test, train, ref}

\begin{table}[t]
\centering\footnotesize
\caption{Average per-benchmark characteristics of the three SPEC CPU2026
workloads (single copy/thread, Sapphire Rapids).}
\label{tab:sizes}
\setlength{\tabcolsep}{5pt}
\begin{tabular*}{\columnwidth}{@{\extracolsep{\fill}}llrrrr}
\toprule
Suite & Size & Instr.\ (G) & IPC & Time (s) & Instr.\ vs.\ ref \\

\midrule
INT Rate (14) & test & 31.8 & 2.49 & 3.2 & 2.3\% \\
 & train & 278.9 & 2.63 & 35.6 & 19.8\% \\
 & ref & 1,411.6 & 2.51 & 154.0 & 100\% \\
\midrule
FP Rate (12) & test & 23.8 & 2.16 & 2.9 & 1.7\% \\
 & train & 221.8 & 2.09 & 30.7 & 15.4\% \\
 & ref & 1,438.4 & 2.23 & 183.8 & 100\% \\
\midrule
INT Speed (13) & test & 36.5 & 2.30 & 4.6 & 0.07\% \\
 & train & 318.1 & 2.29 & 37.9 & 0.65\% \\
 & ref & 49,190.4 & 2.15 & 5,736.4 & 100\% \\
\midrule
FP Speed (13) & test & 66.7 & 2.08 & 8.4 & 0.18\% \\
 & train & 2,202.7 & 2.01 & 195.1 & 6.0\% \\
 & ref & 36,557.9 & 1.82 & 5,633.3 & 100\% \\
\bottomrule
\end{tabular*}
\end{table}

Besides ref, SPEC provides the smaller test and train workloads (Section~\ref{sec:bg-spec}). Because ref runs are long, simulation studies sometimes run train or test in full instead. Table~\ref{tab:sizes} shows how much smaller they are. For the Rate suites, train executes 15--20\% of the ref instruction count on average and test about 2\%. For the Speed suites the gap is far larger: train executes 0.65--6\% of ref and test less than 0.2\%.

\begin{figure*}[t]
\centering
\includegraphics[width=\textwidth]{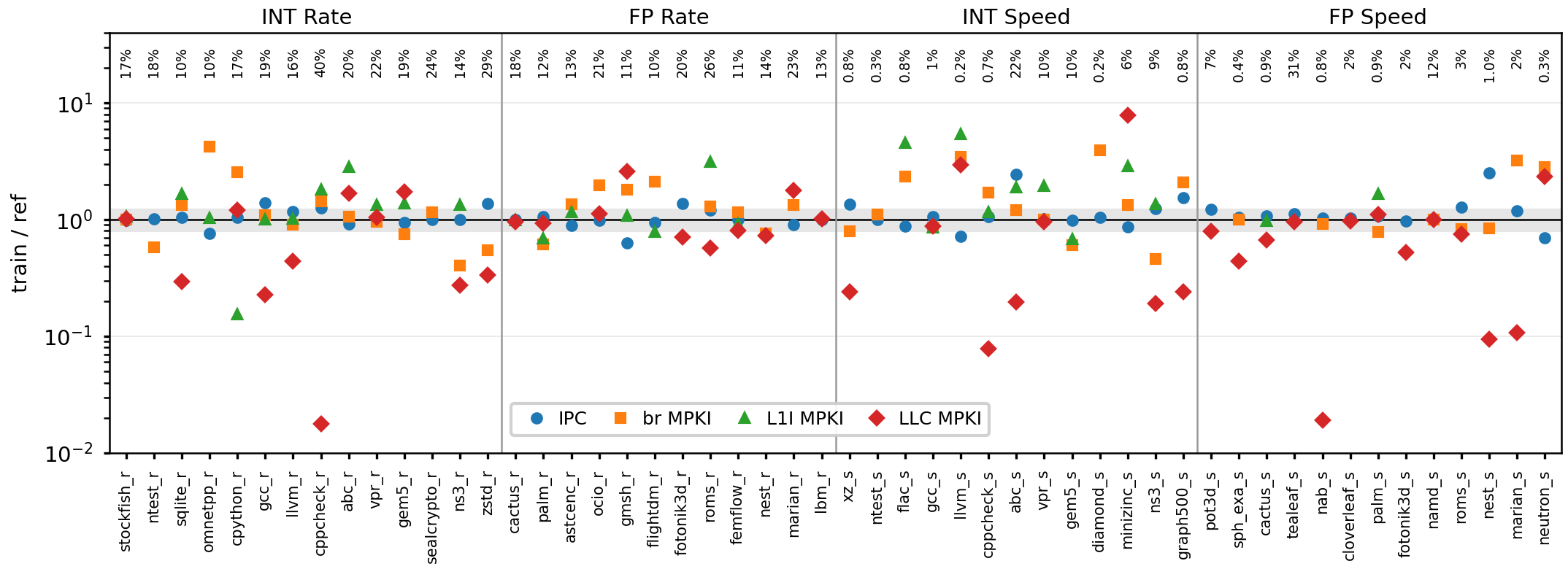}
\caption{Ratio of train to ref for IPC and three MPKIs, per benchmark (log scale; shading marks $\pm$20\%). Numbers at the top give the train instruction count as a percentage of ref. MPKI ratios are omitted where the ref value is below 0.05.}
\label{fig:sizes}
\end{figure*}

The suite averages suggest that the smaller workloads behave like ref: within every suite, the mean IPC of the three sizes differs by less than 15\%. Per benchmark, the picture is different (Figure~\ref{fig:sizes}). The train IPC stays within 20\% of ref for 36 of the 52 benchmarks, but cache behavior can differ sharply even when IPC agrees: \texttt{714.cpython\_r} runs within 5\% of its ref IPC under train, yet its L1I MPKI drops to 15\% of ref, and \texttt{708.sqlite\_r} runs at about the same IPC while its LLC MPKI drops to 29\%. Branch behavior changes as well: \texttt{710.omnetpp\_r} has four times the branch MPKI under train. The Speed train workloads shrink the working set most: in nine Speed benchmarks the LLC MPKI falls below half of ref, and two of them, \texttt{829.abc\_s} and \texttt{867.nest\_s}, run at 2.4--2.5 times their ref IPC. Only a few benchmarks, such as \texttt{706.stockfish\_r} and \texttt{857.namd\_s}, behave alike under both workloads in every metric.

Train can therefore stand in for ref when only a rough IPC is needed, but not in studies of the memory hierarchy or the branch predictor, and test is too small to be representative at all. Rather than running a smaller workload in full, Section~\ref{sec:sampling} examines how well ref itself can be sampled.

\subsection{Compiler Sensitivity}
\label{sec:compiler}
Characterization results are usually reported for a single compiler version. To see how much that choice matters, we rebuilt all four suites with GCC 13.3.1 and repeated the whole-run measurements on the same machine with the same flags; the rest of this paper uses GCC 15.2.0. Table~\ref{tab:compiler} lists the benchmarks whose instruction count or IPC changes by more than 10\%, giving the ratio of the GCC 15 result to the GCC 13 result for instruction count, IPC, and run time.

For most benchmarks the compiler version matters little: 43 of the 52 benchmarks change both their instruction count and their IPC by less than 10\%. The other nine change by more, some by several times, and in both directions. \texttt{706.stockfish\_r} executes only 21\% as many instructions under GCC 15 and its IPC halves, yet it finishes in 42\% of the time. \texttt{838.diamond\_s} keeps nearly the same instruction count but raises its IPC by 1.87$\times$ and finishes in 58\% of the time. \texttt{854.graph500\_s} moves the other way: with an unchanged instruction count, its IPC drops to 0.79 of the GCC 13 value and the run takes 26\% longer. In \texttt{749.fotonik3d\_r} and \texttt{820.cloverleaf\_s}, GCC 15 removes 16--18\% of the instructions, but IPC falls by 14--15\%, so the run time changes by less than 4\%.

IPC is therefore a property of a benchmark and its compiler together: IPC values from studies that use different compilers are not directly comparable, and when the instruction count changes, a lower IPC does not mean a slower run.

\begin{table}[!t]
\centering\footnotesize
\caption{Ratio of GCC 15 to GCC 13 (changes above 10\%).}
\label{tab:compiler}
\begin{tabular*}{\columnwidth}{@{\extracolsep{\fill}}lrrr}
\toprule
Benchmark & Instr. & IPC & Run time \\
\midrule
\texttt{706.stockfish\_r}  & 0.208 & 0.501 & 0.420 \\
\texttt{749.fotonik3d\_r}  & 0.816 & 0.847 & 0.964 \\
\texttt{766.femflow\_r}    & 0.887 & 1.104 & 0.805 \\
\texttt{772.marian\_r}     & 0.777 & 1.150 & 0.677 \\
\texttt{838.diamond\_s}    & 1.073 & 1.867 & 0.580 \\
\texttt{854.graph500\_s}   & 0.997 & 0.791 & 1.259 \\
\texttt{820.cloverleaf\_s} & 0.840 & 0.862 & 0.966 \\
\texttt{822.palm\_s}       & 1.104 & 1.062 & 1.037 \\
\texttt{872.marian\_s}     & 0.774 & 0.950 & 0.815 \\
\bottomrule
\end{tabular*}
\end{table}

\subsection{Cross-Platform Comparison}
\label{sec:cross}

\begin{figure*}[t]
\centering
\includegraphics[width=\textwidth]{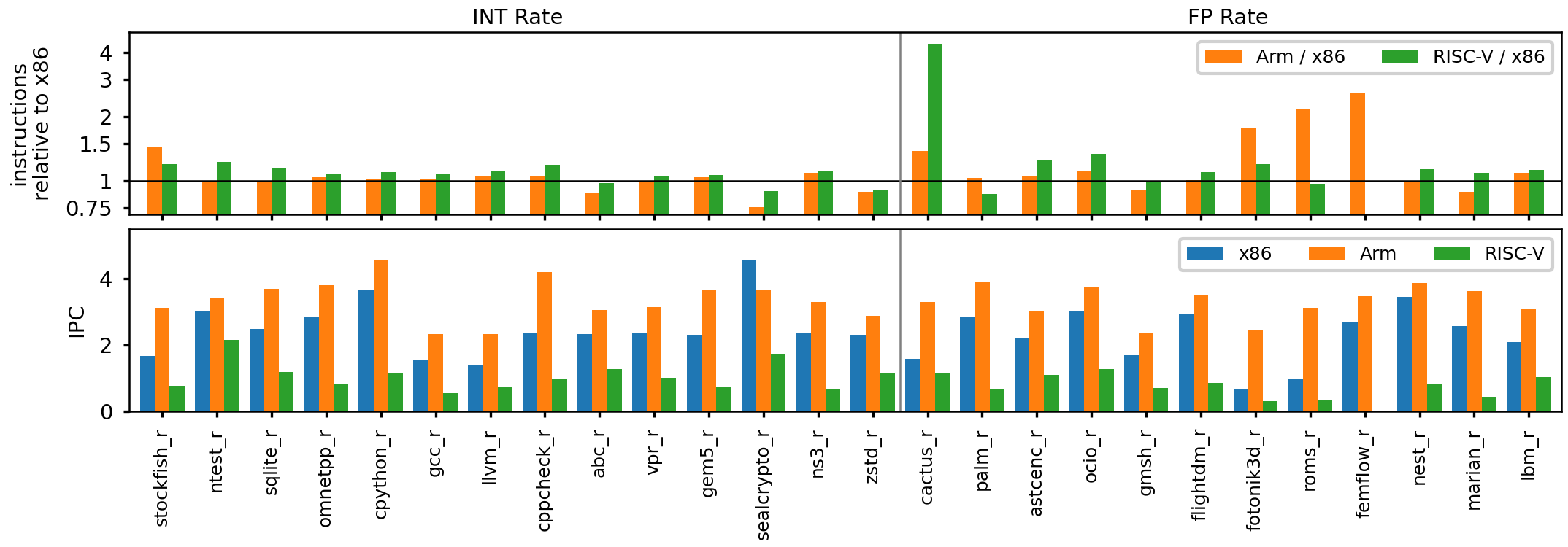}
\caption{Rate suites on three platforms. Top: dynamic instruction count of the Arm and RISC-V builds relative to x86 (log scale). Bottom: IPC on x86 (Sapphire Rapids), Arm (Neoverse V2), and RISC-V (SpacemiT X100). \texttt{766.femflow\_r} is missing on RISC-V.}
\label{fig:cross}
\end{figure*}

Figure~\ref{fig:cross} compares the Rate suites on the three platforms of Table~\ref{tab:platforms}. We compare only instruction count and IPC, which all platforms define identically; the cache and branch events that \texttt{perf} exposes under common names count different things on each vendor's hardware. On the integer suite the three ISAs execute nearly the same number of instructions: the geometric-mean ratio to x86 is 1.01 for Arm and 1.08 for RISC-V, with \texttt{750.sealcrypto\_r} (0.75 on Arm) and \texttt{706.stockfish\_r} (1.45 on Arm) as the largest deviations. The floating-point suite diverges more: Arm and RISC-V both execute 25\% more instructions than x86 on average, up to 2.58$\times$ for \texttt{766.femflow\_r} on Arm and 4.42$\times$ for \texttt{709.cactus\_r} on RISC-V. Per-instruction metrics are therefore comparable across ISAs for most integer benchmarks, but not for floating-point ones, where the same work takes a different number of instructions on each platform.

IPC is highest on Arm and lowest on RISC-V for every benchmark except \texttt{750.sealcrypto\_r}, where x86 leads; the geometric means are 2.40, 3.32, and 1.00 on x86, Arm, and RISC-V for the integer suite and 2.02, 3.26, and 0.71 for floating point. Since the platforms differ in both clock frequency and instruction count, IPC measures work per cycle rather than performance.

\subsection{Bottleneck Classification}
\label{sec:bottleneck}

\begin{figure*}[t]
\centering
\includegraphics[width=\textwidth]{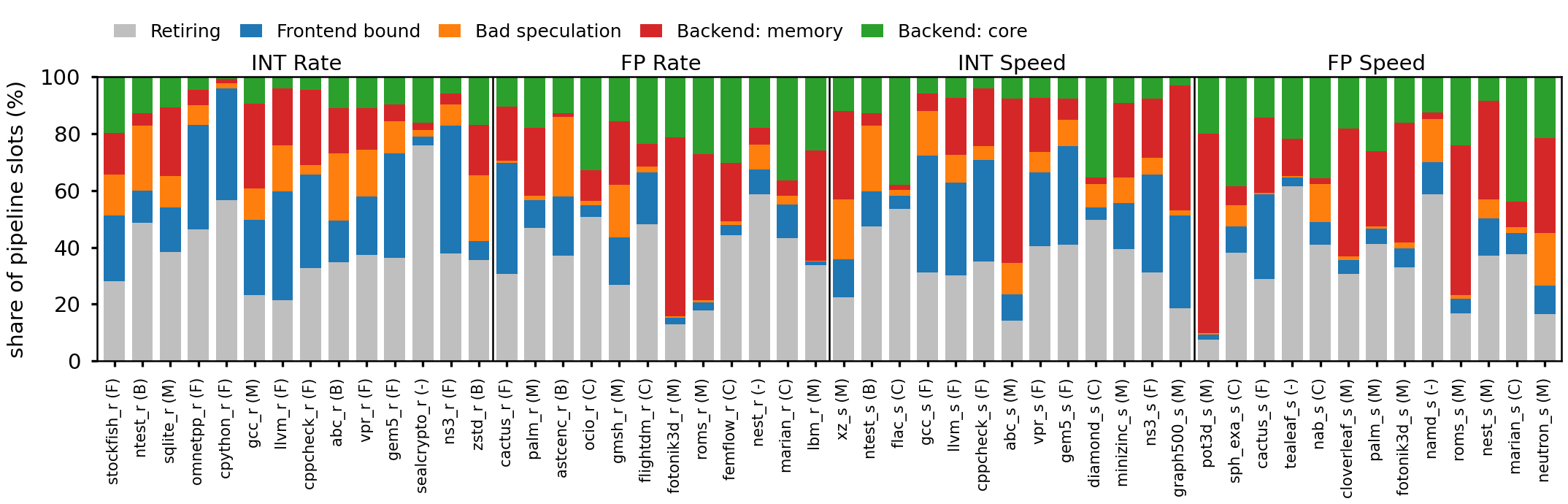}
\caption{Top-down breakdown of pipeline slots for all 52 benchmarks. The letter after each name gives the dominant bottleneck: F frontend, B bad speculation, M memory, C core, and -- balanced.}
\label{fig:tma}
\end{figure*}

To summarize where each benchmark spends its pipeline slots on x86, we split the backend-bound share of Table~\ref{tab:wholerun} into its two level-2 components, which call for different remedies: memory bound, where the backend waits for data from the memory hierarchy, and core bound, where it is limited by execution resources or dependences. Each benchmark is assigned to the largest of four stall categories: frontend, bad speculation, memory, and core. A benchmark that retires useful work in at least half of its slots and loses less than a quarter of them to any single category is labeled balanced. Figure~\ref{fig:tma} shows the breakdown and class of every benchmark.

The integer and floating-point suites have different bottlenecks. Frontend stalls dominate 14 of the 27 integer benchmarks, led by \texttt{753.ns3\_r} (45\% of slots) and \texttt{821.gcc\_s} (41\%). A high frontend share does not always come from L1I misses: \texttt{727.cppcheck\_r} and \texttt{854.graph500\_s} spend about a third of their slots frontend-bound with an L1I MPKI below 1. Memory stalls dominate 12 of the 25 floating-point benchmarks, led by \texttt{800.pot3d\_s} (70\%) and \texttt{749.fotonik3d\_r} (63\%). Core stalls dominate nine benchmarks, seven of them floating point; they include \texttt{872.marian\_s} (44\%) and \texttt{838.diamond\_s} (35\%), the latter with almost no cache misses. Bad speculation dominates five benchmarks, and four are balanced, such as \texttt{750.sealcrypto\_r} and \texttt{811.tealeaf\_s}, which retire useful work in 76\% and 62\% of their slots.

Of the 14 benchmarks that appear in both a Rate and a Speed version, 11 fall into the same class in both. The exceptions are \texttt{gcc}, which moves from memory bound in Rate to frontend bound in Speed; \texttt{abc}, from bad speculation to memory bound; and \texttt{nest}, from balanced to memory bound. The classes offer a way to choose benchmarks by the structure they stress, but like every number in this section, they summarize entire runs that last minutes to hours. These averages hide substantial variation, both across the inputs of a benchmark and over time within a single input (Section~\ref{sec:phase}).

%% file: sections/5_phase.tex
\section{Phase-Level Characterization}
\label{sec:phase}

This section examines how behavior changes over time within each run, using the 10\,ms interval series of the 90 SPECrate inputs (57 integer, 33 floating point) on the x86 platform (Section~\ref{sec:method-hw}). An interval is included only if its cycle count shows that the program occupied the core for at least half of it, which excludes partially idle intervals. We call an interval \emph{slow} if its IPC is below 0.3.

\subsection{Five Patterns of Time-Varying Behavior}
\label{sec:phase-patterns}
\begin{figure}[!t]
\centering
\includegraphics[width=\columnwidth]{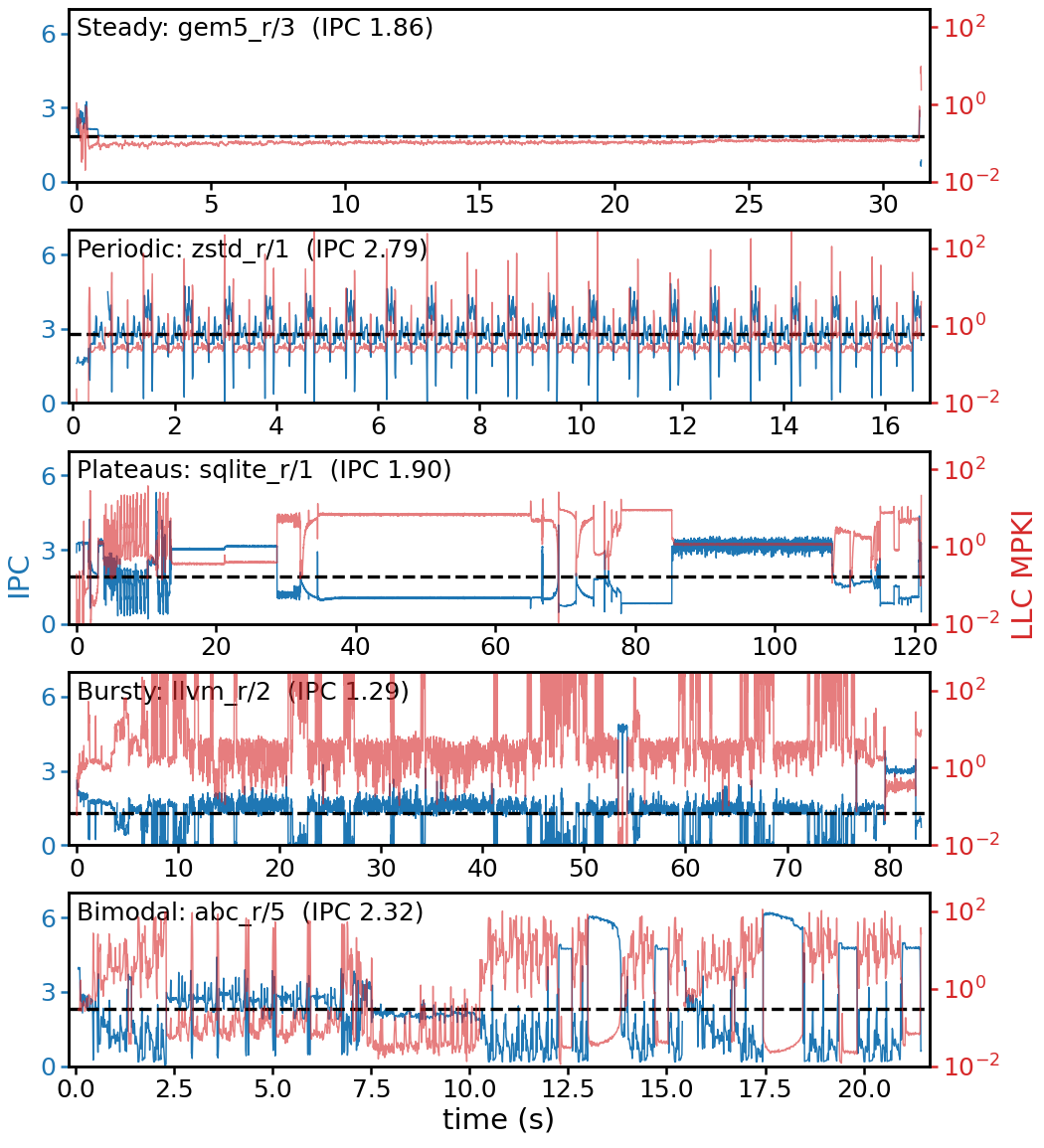}
\caption{Five patterns of time-varying behavior. Blue: IPC per 10\,ms interval; red: LLC MPKI (log scale); dashed: whole-run IPC.}
\label{fig:patterns}
\end{figure}

Figure~\ref{fig:patterns} shows one run for each of the five patterns we observe. \texttt{735.gem5\_r/3} is \emph{steady}: its IPC stays at 1.83--1.87 for nearly the whole run, and the whole-run average describes it well. \texttt{777.zstd\_r/1} is \emph{periodic}: a short IPC pattern repeats at regular intervals throughout the run. \texttt{708.sqlite\_r/1} moves through \emph{plateaus}, segments of tens of seconds at distinct IPC levels, such as about 3 in one and about 1.0 in another; its whole-run IPC of 1.90 matches none of them. \texttt{723.llvm\_r/2} is \emph{bursty}: its IPC stays near 1.5 but drops to almost zero in short bursts throughout the run. \texttt{729.abc\_r/5} is \emph{bimodal}: in the second half of its run, it alternates between segments at an IPC of 5--6 and segments below 1. In \texttt{sqlite\_r/1}, \texttt{llvm\_r/2}, and \texttt{abc\_r/5}, the low-IPC segments coincide with high LLC MPKI.

\subsection{Distribution of IPC Within and Across Inputs}
\label{sec:phase-dist}

\begin{figure*}[!t]
\centering
\includegraphics[width=\textwidth]{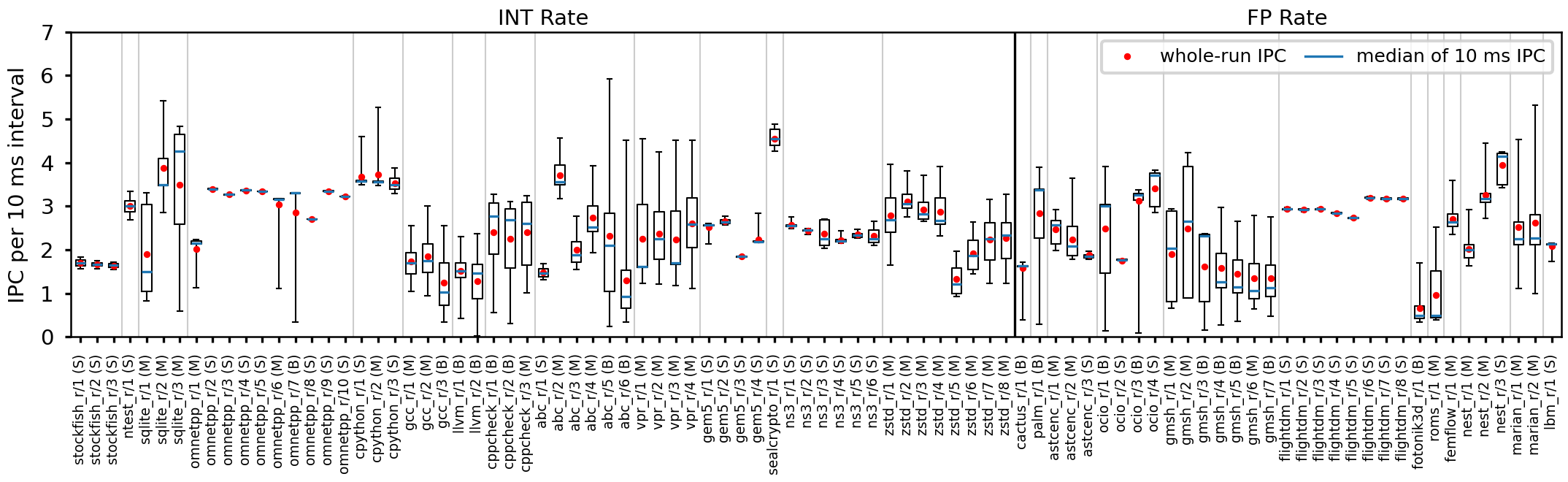}
\caption{Per-interval IPC of the 90 SPECrate inputs (10\,ms intervals). Boxes span the 25th to 75th percentiles with a line at the median; whiskers span the 5th to 95th percentiles; red dots mark the whole-run IPC. Letters after each name give the phase class: S steady, M multi-phase, B bursty.}
\label{fig:box}
\end{figure*}

Figure~\ref{fig:box} shows the distribution of per-interval IPC for all 90 inputs, together with the whole-run IPC of each input. The spread within an input is often large: the 95th-percentile IPC is at least twice the 5th percentile for 40 inputs (24 integer, 16 floating point) and at least ten times for nine. The largest ratios belong to \texttt{723.llvm\_r/2}, whose 5th-percentile IPC is only 0.02 against 2.37 at the 95th, and to \texttt{736.ocio\_r/3} (0.10 against 3.38). \texttt{729.abc\_r/5} spans 0.24 to 5.93, a wider range than the whole-run IPCs of all 90 inputs, which lie between 0.66 and 4.56.

The whole-run IPC, the single number that program-level studies report, can sit far from where the program spends most of its time. For 23 inputs it differs from the median interval IPC by more than 10\%, in both directions: it is 94\% above the median for \texttt{765.roms\_r/1}, 39\% above for \texttt{734.vpr\_r/1} and \texttt{729.abc\_r/6}, and 31\% below for \texttt{737.gmsh\_r/3}. Inputs of the same benchmark can also differ widely. The whole-run IPCs of the six \texttt{abc\_r} inputs range from 1.29 to 3.72, and those of \texttt{zstd\_r}, \texttt{sqlite\_r}, \texttt{nest\_r}, and \texttt{ocio\_r} differ by a factor of about 2, whereas the inputs of \texttt{stockfish\_r}, \texttt{cpython\_r}, \texttt{cppcheck\_r}, and \texttt{marian\_r} stay within 7\% of each other.

\subsection{Classification}
\label{sec:phase-class}

We classify each input by two statistics of its interval IPC. An input is \emph{steady} (S) if its 95th-percentile IPC is at most 1.3 times its 5th percentile, that is, if the middle 90\% of its intervals stay within 30\% of each other. Among the remaining inputs, an input is \emph{bursty} (B) if slow intervals take at least 2\% of its run time, and \emph{multi-phase} (M) otherwise. In Figure~\ref{fig:patterns}, \texttt{gem5\_r/3} is steady, \texttt{zstd\_r/1} and \texttt{sqlite\_r/1} are multi-phase, and \texttt{llvm\_r/2} and \texttt{abc\_r/5} are bursty. Of the 90 inputs, 38 are steady, 35 multi-phase, and 17 bursty; the floating-point suite has a larger share of bursty inputs (9 of 33) than the integer suite (8 of 57).

The classification has to be made per input rather than per benchmark. Seventeen of the 26 benchmarks place all their inputs in one class, for example all eight \texttt{flightdm\_r} inputs are steady and all four \texttt{vpr\_r} inputs multi-phase. The other nine mix classes: the inputs of \texttt{omnetpp\_r} and \texttt{abc\_r} span all three; those of \texttt{cpython\_r}, \texttt{astcenc\_r}, and \texttt{nest\_r} are steady or multi-phase; those of \texttt{gcc\_r}, \texttt{cppcheck\_r}, and \texttt{gmsh\_r} are multi-phase or bursty; and those of \texttt{ocio\_r} are steady or bursty.

The three thresholds are heuristic, so we varied each one separately: the steady threshold to 1.2 and 1.5, the slow-interval IPC to 0.2 and 0.5, and the bursty time share to 1\% and 3\%. No single change moves more than 10 of the 90 inputs to another class, and raising the slow-interval threshold to 0.5 moves only two.

\subsection{Time Share versus Instruction Share}
\label{sec:phase-share}

\begin{figure}[t]
\centering
\includegraphics[width=\columnwidth]{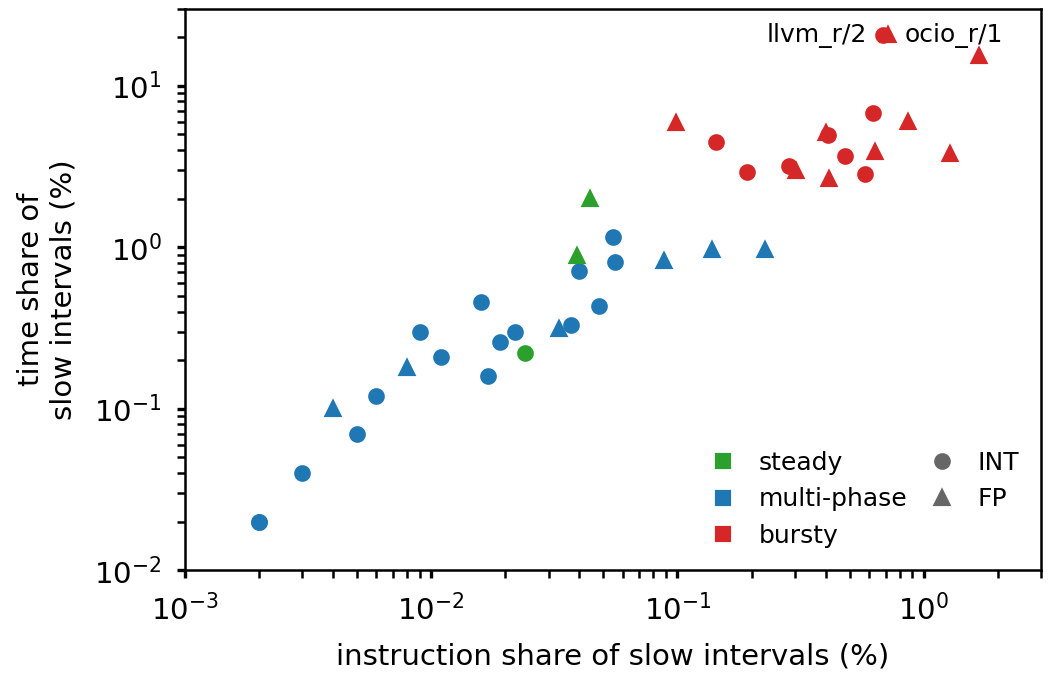}
\caption{Share of run time versus share of instructions in slow intervals (IPC $<$ 0.3), colored by phase class (log scales).}
\label{fig:ext}
\end{figure}

Slow intervals weigh far more in run time than in instructions. Figure~\ref{fig:ext} compares, for the inputs with slow intervals, the share of run time they take with the share of instructions they execute. Slow intervals take at least 1\% of the run time in 19 of the 90 inputs (9 integer, 10 floating point). Because a slow interval executes few instructions, its time share far exceeds its instruction share: for these 19 inputs the ratio ranges from 3.0 to 59. In \texttt{723.llvm\_r/2} and \texttt{736.ocio\_r/1}, slow intervals take about 20\% of the run time but execute only 0.7\% of the instructions.

This gap matters for sampled simulation, which divides a run by instruction count and weights each sample by the instructions it covers. In these two inputs, a fifth of the run time falls on less than 1\% of the instructions; a sample that misrepresents that 1\% misses a fifth of the cycles. Section~\ref{sec:sampling} examines how this affects SimPoint.

%% file: sections/6_sampling.tex
\section{Sampling}
\label{sec:sampling}
This section evaluates SimPoint on the 90 inputs with hardware measurements as the ground truth. For each input we record the cycles of every 100-million-instruction interval and align these intervals with the BBVs that SimPoint clusters (Section~\ref{sec:method}), so the measured CPI of every interval SimPoint might select is known. The CPI that SimPoint predicts is the weighted sum of the measured CPIs of its representatives, and its error is relative to the measured whole-run CPI.

\subsection{Interval Size}
\label{sec:sampling-interval}
\begin{figure}[t]
\centering
\includegraphics[width=\columnwidth]{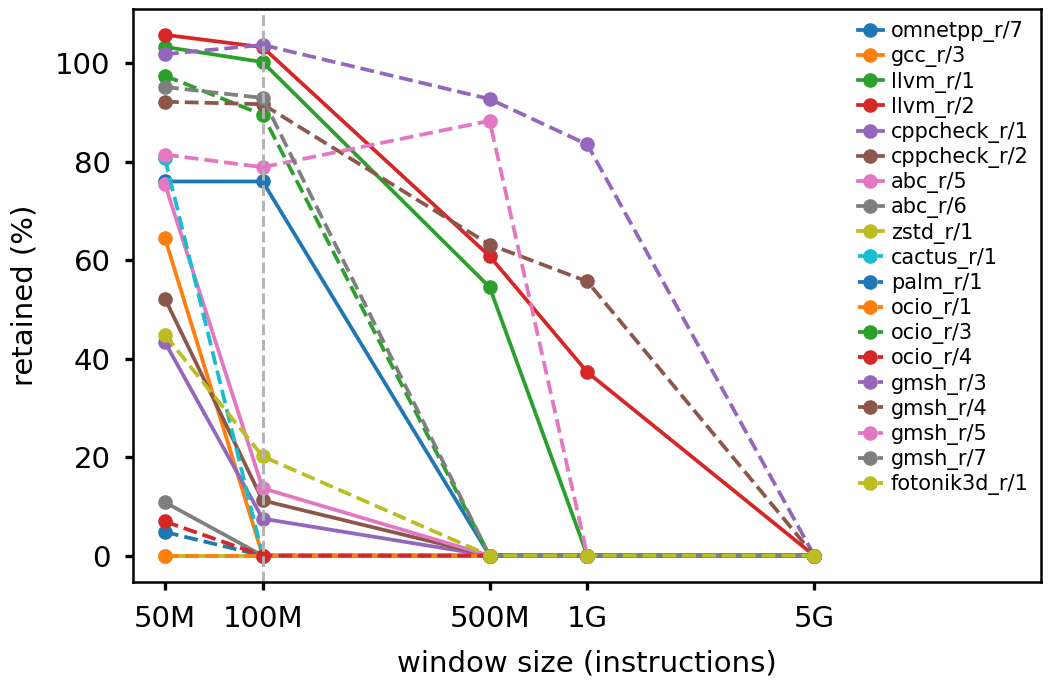}
\caption{Slow-interval time share retained at each window size, relative to its 10\,ms value; solid lines are integer and dashed lines floating-point inputs.}
\label{fig:retention}
\end{figure}

SimPoint divides a run into intervals of a fixed number of instructions, 100 million by default, and we keep that size. An interval this large hides most of the slow intervals. Figure~\ref{fig:retention} regroups the 10\,ms measurements into windows of 50 million to 5 billion instructions and shows each input's slow-interval time share relative to its 10\,ms value, for the 19 inputs in which slow intervals take at least 1\% of the run time. At 100 million instructions, ten of the 19 inputs retain at most 14\% of their slow time share and eight retain more than 75\%, among them the two \texttt{llvm\_r} and four \texttt{gmsh\_r} inputs. At 500 million instructions only five inputs retain any slow windows, and at 5 billion none. Short slow bursts are therefore not separate intervals for SimPoint but parts of ordinary ones.

\subsection{SimPoint Accuracy}
\label{sec:sampling-accuracy}

\begin{figure*}[!t]
\centering
\includegraphics[width=\textwidth]{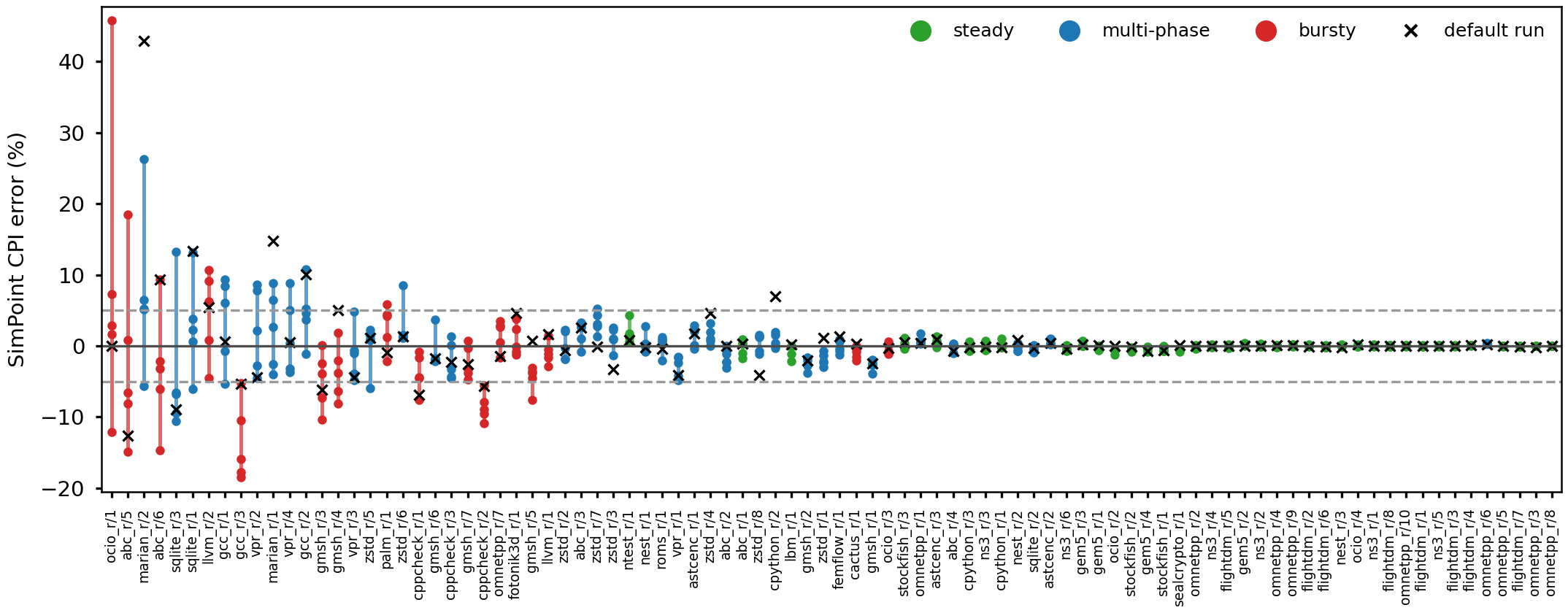}
\caption{SimPoint CPI error for every input: the default run (black cross) and five random seeds (dots, joined by their range), sorted by seed range.}
\label{fig:seeds}
\end{figure*}

With its default settings, SimPoint picks the number of clusters for each input automatically (between 3 and 30 across our inputs, median 18) and predicts the CPI within 2\% for 64 of the 90 inputs and within 5\% for 76. Five inputs are off by more than 10\%: \texttt{772.marian\_r/2} (+42.9\%), \texttt{772.marian\_r/1} (+14.7\%), \texttt{708.sqlite\_r/1} (+13.3\%), \texttt{729.abc\_r/5} ($-$12.6\%), and \texttt{721.gcc\_r/2} (+10.1\%). Figure~\ref{fig:seeds} shows the error of every input as a black cross. Accuracy follows the phase class of Section~\ref{sec:phase-class}: all 38 steady inputs are within 0.9\%, and the 14 inputs with errors of 5\% or more are all multi-phase or bursty, namely 8 of the 17 bursty inputs and 6 of the 35 multi-phase ones. The two suites behave alike: 25 of the 33 floating-point inputs are within 2\%, as are 39 of the 57 integer inputs. The errors have no consistent sign: of the 26 inputs with errors of at least 2\%, SimPoint overestimates the CPI for 11 and underestimates it for 15.

\subsection{Sensitivity to the Random Seed and to K}
\label{sec:sampling-seed}

SimPoint's clustering starts from a random projection and random initial centroids, so a different seed yields a different set of clusters and representatives. We repeated the clustering of every input with five seeds; Figure~\ref{fig:seeds} shows the range of the resulting errors. For the 38 steady inputs the range never exceeds 3.6 percentage points and stays below 2 points for 35 of them. For the other classes it is wide: 13 of the 35 multi-phase inputs and 12 of the 17 bursty inputs have a range above 5 points, the median range is 3.5 points for multi-phase and 6.7 points for bursty inputs, and the widest, \texttt{736.ocio\_r/1}, spans 58 points, from $-$12\% to +46\%, although its default run is within 0.1\%. For these inputs a single SimPoint run is one draw from a wide distribution. The default run can also fall outside the range of the five seeds: \texttt{772.marian\_r/2} is off by +42.9\% where the seeds span $-$5.7\% to +26.2\%.

Fixing $K$ at its maximum of 30, instead of letting SimPoint choose it, does not remove the problem: the mean absolute error over the 90 inputs falls from 2.5\% to 2.3\%, but 9 inputs get worse by more than one point, \texttt{708.sqlite\_r/3} still exceeds 10\% ($-$10.1\%), and \texttt{736.ocio\_r/2}, within 0.1\% in the default run, is off by 72\%.

\subsection{Where the Error Comes From}
\label{sec:sampling-source}

SimPoint can err in two ways: its clusters can group intervals that behave differently, or the representative it picks for a cluster can be atypical. The measured CPI of every interval lets us separate the two. We keep SimPoint's default clusters and weights but replace every representative by the member of its cluster whose CPI is closest to the cluster's cycle-weighted CPI, an oracle that uses the measured answer. This oracle brings every input within 2.6\% of the true CPI, and 88 of the 90 within 2\%. The default clusters are therefore adequate; the error comes from the choice of representatives.

SimPoint picks the member nearest the cluster centroid in BBV space, and nearness in code does not imply typical speed. To see how much the choice matters, we replaced each representative by a random member of its cluster, 200 times per input. For \texttt{729.abc\_r/5} the 5th to 95th percentiles of the resulting errors span $-$19\% to +55\%, and for \texttt{772.marian\_r/2} $-$15\% to +30\%, whereas for every steady input they span at most 4.6 points. 

\begin{figure}[t]
\centering
\includegraphics[width=\columnwidth]{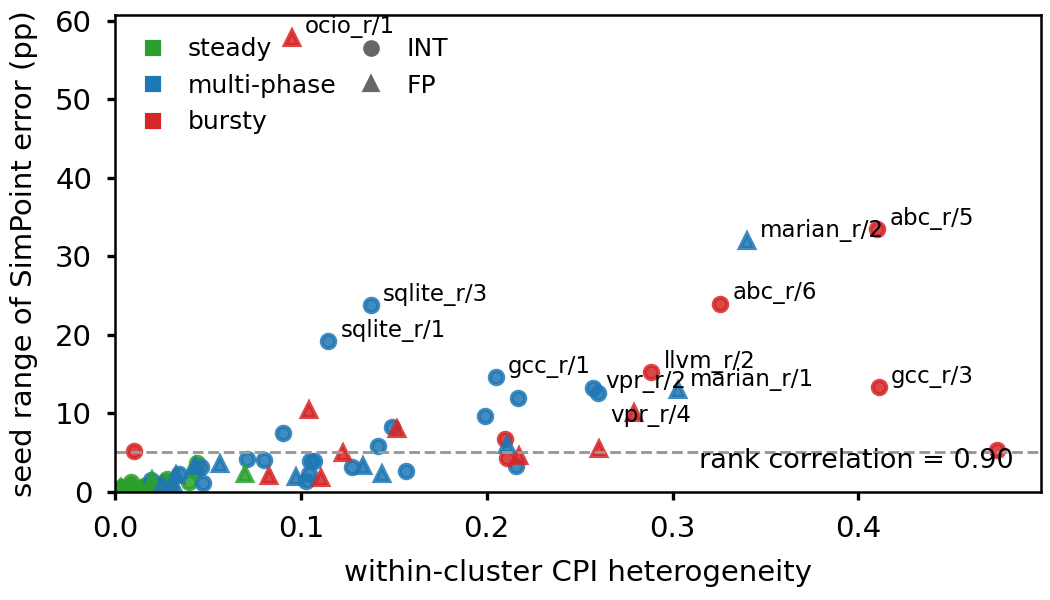}
\caption{Within-cluster CPI heterogeneity versus the seed range of the SimPoint error, one point per input.}
\label{fig:hetero}
\end{figure}

How much the choice matters depends on how much CPI varies among the members of a cluster, which we measure as the coefficient of variation of the member CPIs, averaged over clusters by weight. This heterogeneity has a rank correlation of 0.90 with the seed range, the spread of the SimPoint error over five seeds (Figure~\ref{fig:hetero}), and it follows the phase classes: its median is 0.008 for steady, 0.107 for multi-phase, and 0.211 for bursty inputs. Low heterogeneity is not a guarantee: \texttt{714.cpython\_r/2} has a heterogeneity of 0.035, yet its default run is off by +7.0\%, because the representative of a cluster that holds 36\% of the run has an IPC of 3.02 where the cluster averages 3.57. A BBV records only which code ran, so it cannot separate intervals that execute the same code at different speeds.

\subsection{Recommendations}
\label{sec:sampling-rec}

Our results suggest three practices for sampling CPU2026 with SimPoint.

\begin{enumerate}
\item \emph{Classify the input first.} A single native run with 10\,ms counters, as in Section~\ref{sec:phase}, takes no longer than the program itself and determines the phase class. For the 38 steady inputs, default SimPoint was within 0.9\% of the measured CPI and varied by at most 3.6 percentage points across seeds; for such inputs one run suffices.
\item \emph{For multi-phase and bursty inputs, run several seeds and report the range.} Across five seeds, 25 of these 52 inputs vary by more than 5 percentage points and \texttt{736.ocio\_r/1} by 58, and the default run can fall near either end of the range or outside it (Figure~\ref{fig:seeds}). A single run is one sample from this range.
\item \emph{Do not rely on a larger $K$.} Fixing $K$ at 30 lowered the mean error from 2.5\% to 2.3\%, but made nine inputs worse and left \texttt{708.sqlite\_r/3} at $-$10.1\% and \texttt{736.ocio\_r/2} at +72\%.
\end{enumerate}

%% file: sections/7_conclusion.tex
\section{Conclusion}
\label{sec:conclusion}

In this paper, we characterized SPEC CPU2026 at three levels on real hardware. At the program level, we reported whole-run metrics for all four suites, classified the benchmarks by their dominant bottleneck using top-down analysis, and examined how the results change with the workload size, the compiler version, and the platform, comparing x86, Arm, and RISC-V. At the phase level, we measured how performance varies within each run of the SPECrate inputs, identified recurring patterns of time-varying behavior, and classified the inputs as steady, multi-phase, or bursty. The whole-run average describes the steady inputs well but hides much of the behavior of the others, and slow phases weigh far more in run time than in instruction count. At the sample level, we validated SimPoint against hardware measurements of every interval it can select. SimPoint is accurate on steady inputs but unreliable on inputs with phase structure, where its result also depends on the random seed; the error comes not from how SimPoint groups intervals but from which interval it picks to represent each group. Based on these findings, we gave practical recommendations for sampling CPU2026 with SimPoint. As CPU2026 becomes the standard benchmark suite for evaluating new processors, we expect these results to offer valuable insights to the computer architecture and performance evaluation communities.

%% file: sections/8_acknowledgement.tex
\section*{Acknowledgment}
\label{sec:acknowledgment}
This work was performed under the auspices of the U.S. Department of Energy by Lawrence Livermore National Laboratory under Contract DE-AC52-07NA27344. LLNL-CONF-2026503. This paper describes objective technical results and analysis. Any subjective views or opinions that might be expressed in the paper do not necessarily represent the views of the U.S. Department of Energy or the United States Government.

%% file: refs.bib
@article{henning2000cpu2000,
  author  = {Henning, John L.},
  title   = {{SPEC CPU2000}: Measuring {CPU} Performance in the New Millennium},
  journal = {Computer},
  volume  = {33},
  number  = {7},
  pages   = {28--35},
  month   = jul,
  year    = {2000},
  doi     = {10.1109/2.869367}
}

@article{henning2006cpu2006,
  author  = {Henning, John L.},
  title   = {{SPEC CPU2006} Benchmark Descriptions},
  journal = {ACM SIGARCH Computer Architecture News},
  volume  = {34},
  number  = {4},
  pages   = {1--17},
  month   = sep,
  year    = {2006},
  doi     = {10.1145/1186736.1186737}
}

@inproceedings{bucek2018cpu2017,
  author    = {Bucek, James and Lange, Klaus-Dieter and von Kistowski, J{\'o}akim},
  title     = {{SPEC CPU2017}: Next-Generation Compute Benchmark},
  booktitle = {Companion of the ACM/SPEC International Conference on
               Performance Engineering (ICPE)},
  pages     = {41--42},
  year      = {2018},
  doi       = {10.1145/3185768.3185771}
}

@inproceedings{spec2026method,
  author    = {Madhav, Mahesh and Lee, Allen and Mejia, Andres and Moore, Branden
               and Soppadandi, Charan and Cambly, Chris and M{\"u}llner, Christoph
               and Bowers, Daniel and Reiner, David and Bakhvalov, Denis
               and Zhao, Di and Voth, Duane and Xue, Feng
               and Silber-Chaussumier, Fr{\'e}d{\'e}rique and Bucek, James
               and Southern, James and Liu, Jiangning and Himer, Jim
               and Henning, John and Smith, Kevin and Yang, Kristen
               and Kashyap, Kunal and Guy, Mason and Colgrove, Mat
               and Berg, Michael and Battini, Prasad and Joshi, Prasad
               and Prasad, Rohit and Bhattacharya, Shay and Caculo, Sriyash
               and Reimbold, Stefan and Iyengar, Sundar and Smith, Van
               and Todorovski, Zarko},
  title     = {{SPEC CPU}: The Next Generation},
  booktitle = {ACM/IEEE 53rd Annual International Symposium on Computer
               Architecture (ISCA)},
  pages     = {671--687},
  month     = jun,
  year      = {2026},
  doi       = {10.1109/ISCA66397.2026.00059}
}

@inproceedings{citron2003misspeculation,
  author    = {Citron, Daniel},
  title     = {{MisSPECulation}: Partial and Misleading Use of {SPEC CPU2000}
               in Computer Architecture Conferences},
  booktitle = {Proceedings of the 30th Annual International Symposium on
               Computer Architecture (ISCA)},
  pages     = {52--59},
  year      = {2003},
  doi       = {10.1145/859618.859625}
}

@inproceedings{bai2023archexplorer,
  author    = {Bai, Chen and Huang, Jiayi and Wei, Xuechao and Ma, Yuzhe
               and Li, Sicheng and Zheng, Hongzhong and Yu, Bei and Xie, Yuan},
  title     = {{ArchExplorer}: Microarchitecture Exploration via Bottleneck
               Analysis},
  booktitle = {Proceedings of the 56th Annual IEEE/ACM International Symposium
               on Microarchitecture (MICRO)},
  pages     = {268--282},
  year      = {2023},
  doi       = {10.1145/3613424.3614289}
}

@inproceedings{bera2021pythia,
  author    = {Bera, Rahul and Kanellopoulos, Konstantinos and Nori, Anant
               and Shahroodi, Taha and Subramoney, Sreenivas and Mutlu, Onur},
  title     = {{Pythia}: A Customizable Hardware Prefetching Framework Using
               Online Reinforcement Learning},
  booktitle = {Proceedings of the 54th Annual IEEE/ACM International Symposium
               on Microarchitecture (MICRO)},
  pages     = {1121--1137},
  year      = {2021},
  doi       = {10.1145/3466752.3480114}
}

@inproceedings{juffinger2024suit,
  author    = {Juffinger, Jonas and Kalinin, Stepan and Gruss, Daniel
               and Mueller, Frank},
  title     = {{SUIT}: Secure Undervolting with Instruction Traps},
  booktitle = {Proceedings of the 29th ACM International Conference on
               Architectural Support for Programming Languages and Operating
               Systems (ASPLOS), Volume 2},
  pages     = {1128--1145},
  year      = {2024},
  doi       = {10.1145/3620665.3640373}
}

@inproceedings{niederer2025stackengine,
  author    = {Niederer, Silvan and R{\"u}egge, Sandro and Hajiabadi, Ali
               and Razavi, Kaveh},
  title     = {One Flew over the Stack Engine's Nest: Practical
               Microarchitectural Attacks on the Stack Engine},
  booktitle = {Proceedings of the 58th IEEE/ACM International Symposium on
               Microarchitecture (MICRO)},
  pages     = {94--110},
  year      = {2025},
  doi       = {10.1145/3725843.3756058}
}

@misc{spec2026press,
  author       = {{Standard Performance Evaluation Corporation}},
  title        = {{SPEC} Releases the {SPEC CPU} 2026 Benchmark Suites to
                  Address the Latest Advances in {CPU}, Memory, and Compiler
                  Technology},
  howpublished = {Press release},
  month        = may,
  year         = {2026},
  url          = {https://www.spec.org/pressreleases/2026/20260505-spec-releases-cpu-2026-benchmark-suite/}
}

@misc{li2026ut,
  author       = {Li, Ruihao and Jacob, Andrew and Yadwadkar, Neeraja J.
                  and John, Lizy K.},
  title        = {{SPEC CPU2026}: Characterization, Representativeness, and
                  Cross-Suite Comparison},
  howpublished = {arXiv:2605.03713},
  year         = {2026}
}

@misc{kashyap2026amd,
  author       = {Kashyap, Kunal and Ramanathan, Rajiv
                  and Bhattacharya, Shayantika},
  title        = {Performance Characterization of {SPEC CPU} 2026 on {AMD}
                  {EPYC} 9755 Processor},
  howpublished = {arXiv:2609.01527},
  year         = {2026}
}

@misc{alotoom2026fidelity,
  author       = {Al-Otoom, Doa'a and Madhav, Mahesh},
  title        = {Adaptation Fidelity of {SPEC CPU2026}},
  howpublished = {arXiv:2608.27710},
  year         = {2026}
}

@inproceedings{panda2018hpca,
  author    = {Panda, Reena and Song, Shuang and Dean, Joseph and John, Lizy K.},
  title     = {Wait of a Decade: Did {SPEC CPU} 2017 Broaden the Performance
               Horizon?},
  booktitle = {IEEE International Symposium on High Performance Computer
               Architecture (HPCA)},
  pages     = {271--282},
  year      = {2018},
  doi       = {10.1109/HPCA.2018.00032}
}

@inproceedings{limaye2018ispass,
  author    = {Limaye, Ankur and Adegbija, Tosiron},
  title     = {A Workload Characterization of the {SPEC CPU2017} Benchmark
               Suite},
  booktitle = {IEEE International Symposium on Performance Analysis of Systems
               and Software (ISPASS)},
  pages     = {149--158},
  year      = {2018},
  doi       = {10.1109/ISPASS.2018.00028}
}

@article{binkert2011gem5,
  author  = {Binkert, Nathan and Beckmann, Bradford and Black, Gabriel
             and Reinhardt, Steven K. and Saidi, Ali and Basu, Arkaprava
             and Hestness, Joel and Hower, Derek R. and Krishna, Tushar
             and Sardashti, Somayeh and Sen, Rathijit and Sewell, Korey
             and Shoaib, Muhammad and Vaish, Nilay and Hill, Mark D.
             and Wood, David A.},
  title   = {The gem5 Simulator},
  journal = {ACM SIGARCH Computer Architecture News},
  volume  = {39},
  number  = {2},
  pages   = {1--7},
  month   = aug,
  year    = {2011},
  doi     = {10.1145/2024716.2024718}
}

@inproceedings{wunderlich2003smarts,
  author    = {Wunderlich, Roland E. and Wenisch, Thomas F.
               and Falsafi, Babak and Hoe, James C.},
  title     = {{SMARTS}: Accelerating Microarchitecture Simulation via
               Rigorous Statistical Sampling},
  booktitle = {Proceedings of the 30th Annual International Symposium on
               Computer Architecture (ISCA)},
  pages     = {84--95},
  year      = {2003},
  doi       = {10.1109/ISCA.2003.1206991}
}

@inproceedings{sherwood2002simpoint,
  author    = {Sherwood, Timothy and Perelman, Erez and Hamerly, Greg
               and Calder, Brad},
  title     = {Automatically Characterizing Large Scale Program Behavior},
  booktitle = {Proceedings of the 10th International Conference on
               Architectural Support for Programming Languages and Operating
               Systems (ASPLOS)},
  pages     = {45--57},
  year      = {2002},
  doi       = {10.1145/605397.605403}
}

@article{hamerly2005simpoint3,
  author  = {Hamerly, Greg and Perelman, Erez and Lau, Jeremy and Calder, Brad},
  title   = {{SimPoint} 3.0: Faster and More Flexible Program Phase Analysis},
  journal = {Journal of Instruction-Level Parallelism},
  volume  = {7},
  year    = {2005}
}

@inproceedings{lau2005correlation,
  author    = {Lau, Jeremy and Sampson, Jack and Perelman, Erez
               and Hamerly, Greg and Calder, Brad},
  title     = {The Strong Correlation Between Code Signatures and Performance},
  booktitle = {IEEE International Symposium on Performance Analysis of Systems
               and Software (ISPASS)},
  pages     = {236--247},
  year      = {2005},
  doi       = {10.1109/ISPASS.2005.1430578}
}

@inproceedings{yasin2014topdown,
  author    = {Yasin, Ahmad},
  title     = {A Top-Down Method for Performance Analysis and Counters
               Architecture},
  booktitle = {IEEE International Symposium on Performance Analysis of Systems
               and Software (ISPASS)},
  pages     = {35--44},
  year      = {2014},
  doi       = {10.1109/ISPASS.2014.6844459}
}

@inproceedings{singh2019memory,
  author    = {Singh, Sarabjeet and Awasthi, Manu},
  title     = {Memory Centric Characterization and Analysis of {SPEC CPU2017}
               Suite},
  booktitle = {Proceedings of the ACM/SPEC International Conference on
               Performance Engineering (ICPE)},
  pages     = {285--292},
  year      = {2019},
  doi       = {10.1145/3297663.3310311}
}

@article{navarro2019memory,
  author  = {Navarro-Torres, Agust{\'i}n and Alastruey-Bened{\'e}, Jes{\'u}s
             and Ib{\'a}{\~n}ez-Mar{\'i}n, Pablo
             and Vi{\~n}als-Y{\'u}fera, V{\'i}ctor},
  title   = {Memory Hierarchy Characterization of {SPEC CPU2006} and
             {SPEC CPU2017} on the {Intel Xeon Skylake-SP}},
  journal = {PLOS ONE},
  volume  = {14},
  number  = {8},
  pages   = {e0220135},
  year    = {2019},
  doi     = {10.1371/journal.pone.0220135}
}

@inproceedings{hebbar2019,
  author    = {{Hebbar S R}, Ranjan and Milenkovi{\'c}, Aleksandar},
  title     = {{SPEC CPU2017}: Performance, Event, and Energy Characterization
               on the {Core i7-8700K}},
  booktitle = {Proceedings of the ACM/SPEC International Conference on
               Performance Engineering (ICPE)},
  pages     = {111--118},
  year      = {2019},
  doi       = {10.1145/3297663.3310314}
}

@inproceedings{duesterwald2003,
  author    = {Duesterwald, Evelyn and Cascaval, Calin and Dwarkadas, Sandhya},
  title     = {Characterizing and Predicting Program Behavior and its
               Variability},
  booktitle = {International Conference on Parallel Architectures and
               Compilation Techniques (PACT)},
  pages     = {220--231},
  year      = {2003},
  doi       = {10.1109/PACT.2003.1238018}
}

@inproceedings{carlson2014barrierpoint,
  author    = {Carlson, Trevor E. and Heirman, Wim and Van Craeynest, Kenzo
               and Eeckhout, Lieven},
  title     = {{BarrierPoint}: Sampled Simulation of Multi-Threaded
               Applications},
  booktitle = {IEEE International Symposium on Performance Analysis of Systems
               and Software (ISPASS)},
  pages     = {2--12},
  year      = {2014},
  doi       = {10.1109/ISPASS.2014.6844456}
}

@inproceedings{sabu2022looppoint,
  author    = {Sabu, Alen and Patil, Harish and Heirman, Wim
               and Carlson, Trevor E.},
  title     = {{LoopPoint}: Checkpoint-driven Sampled Simulation for
               Multi-threaded Applications},
  booktitle = {IEEE International Symposium on High-Performance Computer
               Architecture (HPCA)},
  pages     = {604--618},
  year      = {2022},
  doi       = {10.1109/HPCA53966.2022.00051}
}

@inproceedings{nair2008cpu2006,
  author    = {Nair, Arun A. and John, Lizy K.},
  title     = {Simulation Points for {SPEC CPU} 2006},
  booktitle = {IEEE International Conference on Computer Design (ICCD)},
  pages     = {397--403},
  year      = {2008},
  doi       = {10.1109/ICCD.2008.4751891}
}

@inproceedings{wu2018hot,
  author    = {Wu, Qinzhe and Flolid, Steven and Song, Shuang and Deng, Junyong
               and John, Lizy K.},
  title     = {Hot Regions in {SPEC CPU2017}},
  booktitle = {IEEE International Symposium on Workload Characterization
               (IISWC)},
  pages     = {71--77},
  year      = {2018},
  doi       = {10.1109/IISWC.2018.8573479}
}

@misc{han2021pinpoints,
  author       = {Han, Haiyang and Hardavellas, Nikos},
  title        = {Public Release and Validation of {SPEC CPU2017} {PinPoints}},
  howpublished = {arXiv:2112.06981},
  year         = {2021}
}

@misc{caculo2025mav,
  author       = {Caculo, Sriyash and Madhav, Mahesh and Baxter, Jeff},
  title        = {Memory Access Vectors: Improving Sampling Fidelity for {CPU}
                  Performance Simulations},
  howpublished = {arXiv:2506.02344},
  year         = {2025}
}

@inproceedings{xue2025simpointplus,
  author    = {Xue, Jiangying and Xiong, Tianyu and Chao, Lingwei
               and Xue, Ruini},
  title     = {{SimPoint+}: More Stable, Accurate and Efficient Program
               Analysis},
  booktitle = {Euro-Par 2025: Parallel Processing},
  publisher = {Springer},
  pages     = {3--17},
  year      = {2025},
  doi       = {10.1007/978-3-031-99857-7_1}
}

@inproceedings{luk2005pin,
  author    = {Luk, Chi-Keung and Cohn, Robert and Muth, Robert and Patil, Harish
               and Klauser, Artur and Lowney, Geoff and Wallace, Steven
               and Reddi, Vijay Janapa and Hazelwood, Kim},
  title     = {{Pin}: Building Customized Program Analysis Tools with Dynamic
               Instrumentation},
  booktitle = {Proceedings of the ACM SIGPLAN Conference on Programming
               Language Design and Implementation (PLDI)},
  pages     = {190--200},
  year      = {2005},
  doi       = {10.1145/1065010.1065034}
}
